\documentclass[prd,twocolumn,showpacs,amsmath]{revtex4}
\usepackage{amssymb,amsmath}
\usepackage{graphicx}
\usepackage{color}

\begin{document}
\title{Quantum Cosmology in $(1+1)$-dimensional Ho\v rava-Lifshitz theory of gravity}

\author{J. P. M. Pitelli}
\email[]{pitelli@ime.unicamp.br}
\affiliation{Departamento de Matem\'atica Aplicada, Universidade Estadual de Campinas, 13083-859, Campinas, SP, Brazil}
\pacs{04.60.Kz, 04.60.Ds}
\begin{abstract}

In a recent paper [Phys. Rev. D 92:084012, 2015], the author studied  the classical $(1+1)$-dimensional Friedmann-Robertson-Walker (FRW) universe filled with a perfect fluid in  Ho\v rava-Lifshitz (HL) theory of gravity. This theory is dynamical due to the anisotropic scaling of space and time. It also resembles the Jackiw-Teitelboim model, in which a dilatonic degree of freedom is necessary for dynamics. In this paper, I will give one step further in the understanding of (1+1)-dimensional HL cosmology by means of the quantization  of the FRW universe filled with a perfect fluid with equation of state (EoS) $p=w\rho$. The fluid will be introduced in the model via Schutz formalism and Dirac's algorithm will be used for quantization. It will be shown that the Schr\"odinger equation for the wave function of the universe has the following properties: for $w=1$ (radiation fluid), the characteristic potential will be exponential, resembling Liouville quantum mechanics; for $w\neq 1$, a characteristic inverse square potential appears in addition to a regular polynomial which depends on the EoS. Explict solutions for a few cases of interesting will be found and the expectation value of the scale factor will be calculated. As in usual quantum cosmology, it will be shown that the quantum theory smooth out the big-bang singularity, but the classical behavior of the universe is recovered  in the low-energy limit.

\end{abstract}

\maketitle

\section {Introduction}

In recent years, Ho\v rava-Lifshitz theory \cite{horava1} has become a valuable alternative theory of gravity. The theory is power countable renormalizable \cite{visser1}  with an anisotropic scaling of the space and time coordinates of the form
\begin{equation}
t\to b^zt, \,\,\,\,\,x^i\to bx^i\,\,\,(i=1,\dots, d), 
\end{equation}
where $z$ denotes the dynamical critical exponent. When $z\neq 1$, the Lorentz symmetry is explicitly broken. However, even not exhibiting relativistic invariance at short distances, General Relativity (GR) is recovered in the low-energy limit.  In cosmological models, HL theory has interesting consequences. It can solve the horizon and flatness problem and can generate scale invariant perturbations for the early universe even without inflation \cite{wang2,kiritsis,mukohyama}. Moreover, due to high orders curvature terms, bouncing and cyclic universes appear as regular solutions \cite{mukohyama2}. 

In $(1+1)$-dimensions, the HL action fails to be a topological invariant, giving rise to a dynamical theory \cite{wang}, unlike Einstein's gravity, which is nondynamical in two-dimensions. Cosmology in $(1+1)$-dimensional HL theory was studied in Ref. \cite{pitelli1}. It was shown that it is very similar to the Jackiw-Teitelboim model \cite{grumiller,cadoni} (where a dilaton scalar field is necessary for dynamics), in the sense that the evolution of the scale factor in both cases are the same. Since two-dimensional HL theory is dynamical, it is amenable to quantization.

Quantum cosmology is based on the Wheeler-DeWitt equation \cite{dewitt}, whose solution is the wave function of the universe as a whole. Generally, the Wheeler-DeWitt equation is very difficult to be handled, since it is defined in the superspace of all possible spatial metrics and all possible field configurations. However, the symmetries of the homogeneous FRW universe allows us to freeze out all but a finite number of degrees of freedom. The remaining degrees of freedom can then be quantized by canonical methods. Another problem with the Wheeler-DeWitt equation is the absence of a natural time. The Hamiltonian constructed using the Arnowitt-Deser-Misner \cite{adm} formalism is a constraint, with no first derivative playing the role of time. This problem is generally circumvented with the introduction of matter (a fluid, for example) in the model. In this way, time is measured by a dynamical degree of freedom related to the matter content. The standard way to introduce a fluid in quantum cosmology is via Schutz formalism \cite{schutz1,schutz2}. When applied to FRW universe, this formalism gives rise to at least one linear conjugated momentum in the Hamiltonian. In this way, a parameter of time is introduced in the model and a Schr\"odinger-like equation can be obtained.

Quantization of the FRW universe filled with a perfect fluid in $(1+3)$-dimensional HL theory was performed in Refs. \cite{pitelli2,vakili}. In 
\cite{pitelli2}, it was shown that one of the main problems of quantum cosmology, namely, the necessity of a boundary condition for the Wheeler-DeWitt equation at the big-bang singularity, is solved in HL theory. The higher order spatial curvature terms in HL quantum cosmology introduce a  repulsive potential near $a=0$ ($a$ is the scale factor)  which shields the classical singularity so that quantum wave packets simply bounce off this barrier. In \cite{vakili}, exact solutions of the Wheeler-DeWitt equation were obtained and the expectation value of the scale factor was analyzed, showing a bouncing behavior near the classical singularity. 

In \cite{wang}, the $(1+1)$-dimensional FRW universe was quantized when only gravity is present and when gravity couples to a scalar field. It was shown that, in both cases, Dirac's algorithm works and normalizable wave functions for some operator orderings were exhibited. In this paper,  the minisuperspace model of FRW universe filled with a perfect fluid is quantized. The Hamiltonian constraint, which leads to the Wheeler-DeWitt equation, was already obtained in \cite{pitelli1}. It depends on the scale factor $a$, its conjugated momentum $p_a$ and the conjugated momentum $p_T$ of a  dynamical degree of freedom of the fluid. However, unlike in standard quantum mechanics, the Hamiltonian constraint in FRW models does not lead to a unique ordering due to the non-commutativity of $\hat{a}$ and $\hat{p}_a$. Different orderings give rise to different transitions between the classical and the quantum theory. By some physical motivation, one must choose a specific operator ordering. Here I show that, with  a particular operator ordering, the Wheeler-DeWitt equation is in Sturm-Liouville form. By performing a Liouville transformation to this equation, an equivalent Schr\"odinger equation is found and its characteristic potential is analyzed. For a radiation fluid with EoS $p=\rho$, the  wave function of the universe  satisfies the free particle Schr\"odinger equation when the cosmological constant $\Lambda$ is null. When $\Lambda\neq 0$,  an exponential (Liouville) potential appears. For $p=w\rho$ and $w\neq 1$,  the Schr\"odinger equation has a characteristic inverse square potential, in addition to a regular polynomial, which depends on the EoS. Unlike the $(1+3)$-dimensional case, there is an infinite number of possible boundary conditions. In order to find explicit solutions for some solvable cases, the Friedrichs boundary condition \cite{reed} is chosen. Then I  compare the expectation values of the scale factor with the classical ones obtained in \cite{pitelli1}.

This paper is organized as follows: in Sec. II,  the Hamiltonian constraint of the $(1+1)$-dimensional FRW universe filled with a perfect fluid will be found. This section is based on Refs. \cite{wang} and \cite{pitelli1}. In Sec. III, I will proceed with the canonical quantization of the model, by finding a Schr\"odinger-like equation which comes from the Hamiltonian constraint by means of a particular choice of operator ordering. By performing a Liouville transformation to this equation, the correspondent Schr\"odinger equation will be found and analyzed. In Sec. IV, some interesting solvable cases will be studied and the evolution of the expectation value of the scale factor will be found. A comparison between the quantum and the classical predictions will be made. Finally, Sec. V contains my final considerations. 

\section{Hamiltonian Constraint}

The anisotropic scaling of space and time in HL theory breaks Lorentz invariance. In this way, the HL action is no longer a topological invariance and the theory is dynamical. Since GR is recovered in the low energy limit, the matter content will be introduced in the model as in General Relativity, i.e., via Schutz formalism of perfect fluids coupled to gravity \cite{schutz1,schutz2}. Then, the total action will be given by $S=S_{HL}+S_f$, in which $S_{HL}$ denotes the action of the FRW minisuperspace in HL theory and $S_f$ denotes the action of the matter content. Given the total action, the Hamiltonian constraint can be obtained by canonical methods. Following Dirac's algorithm of quantization,  we associate to this Hamiltonian constraint an operator which annihilates the wave function of the universe. Due to the matter content, there is a parameter of time in the model and the resulting equation will be a Schr\"odinger-like equation.



In $(1+1)$ dimensions, the HL action is given by \cite{wang}
\begin{equation}
S_{HL}=\int{dtdxN(t)\sqrt{h_{11}}\left(\mathcal{L}_K-\mathcal{L}_V\right)}.
\end{equation}
In the above equation, $N(t)$ is the lapse function, $h_{11}$ is the spatial part of the metric,  $\mathcal{L}_K$ is the kinetic term and $\mathcal{L}_V$ is the potential Lagrangian. The kinetic term depends on the extrinsic curvature $K_{11}$ of the constant leaves $t=\text{constant}$ as
\begin{equation}
\mathcal{L}_K=K^{11}K_{11}-\lambda K^2,
\end{equation}
where $\lambda$ is  a dimensionless constant. Note that, although $\lambda$ breaks Lorentz invariance, the action is still invariant under the group of diffeomorphism of the spatial slices. The potential Lagrangian $\mathcal{L}_V$ is given by
\begin{equation}
\mathcal{L}_V=-2\Lambda-\alpha a^{i}a_{i}, 
\label{potential lagrangian}
\end{equation}
where $\alpha$ is a coupling constant and $a_{i}=\partial_i\ln{N}$.

For the matter content, the action of a perfect fluid coupled to gravity in Schutz formalism \cite{schutz1,schutz2} is given by
\begin{equation}
S_{f}=\int{d^2x\sqrt{-g}p},
\end{equation}
where $p$ is the fluid's pressure, which in turn is related to the fluid's density by the EoS $p=w\rho$. The two-velocity of the fluid depends on its specific enthalpy $\mu$, the specific entropy $S$ and two other potentials $\phi$ and $\theta$ with no clear physical meaning \cite{ahmed}. It is given by
\begin{equation}
U_\nu=\frac{1}{\mu}\left(\phi_{,\nu}+\theta S_{,\nu}\right).
\label{two-velocity}
\end{equation}
By thermodynamical arguments,  Lapchinski and Rubakov \cite{Lapchinskii} found that the pressure depends on the Schutz potentials as
\begin{equation}
p=\frac{w\mu^{1+1/w}}{(1+w)^{1+1/w}}e^{-S/w}.
\end{equation}

Let us consider the FRW metric
\begin{equation}
ds^2=-N(t)^2dt^2+a(t)^2dx^2, 
\end{equation}
where $a(t)$ is the scale factor. In this case, the extrinsic curvature tensor is given by
\begin{equation}
K_{11}=\frac{1}{N}(-\dot{h}_{11}+2\nabla_1N_1)=-\frac{a\dot{a}}{N}.
\label{extrinsic}
\end{equation}
The two-velocity field of the fluid is $U_{\nu}=N\delta^{0}_{\nu}$. By Eq. (\ref{two-velocity}) we have
\begin{equation}
\mu=\left(\frac{\dot{\phi}+\theta \dot{S}}{N}\right).
\label{enthalpy}
\end{equation}

Plugging Eqs. (\ref{extrinsic}) and (\ref{enthalpy}) into the total action $S=S_{HL}+S_{f}$ leads to
\begin{equation}\begin{aligned}
S=&\int{dtdx\left[(1-\lambda)\frac{\dot{a}^2}{Na}+2\Lambda aN\right.}\\&\left.+\frac{w}{(1+w)^{1+1/w}}\left(\frac{\dot{\phi}+\theta\dot{S}}{N}\right)^{1+1/w}Na e^{-S/w}\right].
\end{aligned}
\label{total action}
\end{equation}
Given the total action, we can define the conjugated momenta
\begin{equation}
\left\{\begin{aligned}
&p_a=2(1-\lambda)\frac{\dot{a}}{Na},\\
&p_\phi=\frac{a \mu^{1/w}}{(1+w)^{1/w}}e^{-S/w},\\
&p_{S}=\theta p_{\phi}.
\label{momenta}
\end{aligned}\right.
\end{equation}
The Hamiltonian is then given by
\begin{equation}
\begin{aligned}
H&=p_a\dot{a}+p_\phi\left(\dot{\phi}+\theta\dot{S}\right)-L,\\
&=N\left(\frac{ap_a^2}{4(1-\lambda)}+\frac{p_\phi^{1+w}e^{S}}{a^{w}}-2\Lambda a\right).
\end{aligned}
\label{hamiltonian1}
\end{equation}
Note that the above expression does not contain any linear conjugated momentum. However, if we perform the following canonical transformation
\begin{equation}
\left\{\begin{aligned}
&T=-p_Se^{-S}\phi^{-(1+w)},\\
&p_T=p_{\phi}^{1+w}e^{S},\\
&\bar{\phi}=\phi+(1+w)\frac{p_S}{p_\phi},\\
&\bar{p}_\phi=p_\phi,
\label{canonical transformation}
\end{aligned}\right.
\end{equation}
we arrive at
\begin{equation}
H=N\left(\frac{ap_a^2}{4(1-\lambda)}+\frac{p_T}{a^{w}}-2\Lambda a\right).
\label{hamiltonian2}
\end{equation}
The conjugated momentum $p_T$ appears linearly in the above expression. In this way, $T$ will be a parameter of time after quantization. Hamilton's equations applied to Eq. (\ref{hamiltonian2}) give us
\begin{equation}
\left\{\begin{aligned}
&\dot{p}_T=0\Rightarrow p_T=\text{constant},\\
&\dot{T}=\frac{N}{a^{w}},\\
&\dot{p}_{a}=N\left[-\frac{p_a^2}{4(1-\lambda)}+2\Lambda +w \frac{p_T}{a^{1+w}}\right],\\
&\dot{a}=N\frac{ap_a}{2(1-\lambda)},
\label{equations of motion}
\end{aligned}\right.
\end{equation}
so that $T$ is time in the gauge $N=a^w$.

Finaly, the super-Hamiltonian constraint comes from varying the action with respect to $N$. It is given by
\begin{equation}
\mathcal{H}=\frac{ap_a^2}{4(1-\lambda)}+\frac{p_T}{a^{w}}-2\Lambda a\approx 0.
\end{equation}

\section{Quantization}

To proceed with Dirac's algorithm of quantization of constrained systems \cite{dirac}, we perform the substitutions $p_a\to-i\partial/\partial a$ and $p_T\to -i \partial/\partial T$ and demand that the super-Hamiltonian operator  annihilates the wave function, i.e., $\mathcal{H}\Psi=0$. We still have an operator ordering ambiguity, since $\hat{a}$ and $\hat{p}_a$ do not commute. With the operator ordering $ap_a^2\to \hat{p}_a\hat{a}\hat{p}_a$,  the resulting Schr\"odinger-like equation
\begin{equation}
-\frac{1}{4(1-\lambda)}\frac{\partial}{\partial a}\left(a \frac{\partial \psi(a,T)}{\partial a}\right)-2\Lambda a \psi(a,T)=\frac{i}{a^{w}}\frac{\partial \psi(a,T)}{\partial T}
\label{Sturm-Liouville}
\end{equation}
is already in the Sturm-Liouville form. Note that, in order for the above equation to be formally self-adjoint, the inner product must be given by
\begin{equation}
\left<f(a),g(a)\right>=\int_{0}^{\infty}{f^{\ast}(a)g(a)\frac{da}{a^{w}}}.
\end{equation}

Given Eq. (\ref{Sturm-Liouville}), we can perform a Liouville transformation to obtain an equivalent Schr\"odinger equation 
\begin{equation}
-\frac{\partial^2 y(x,t)}{\partial x^2}+V(x)y(x,t)=i\frac{\partial y(x,t)}{\partial t},
\label{Sch}
\end{equation}
with corresponding inner product 
\begin{equation}
\left<f(x),g(x)\right>=\int{f^{\ast}(x)g(x)dx}.
\end{equation}

This can be done in the following way \cite{everitt}. For a Sturm-Liouville differential equation 

\begin{equation}
-\left(p(a)\phi(a)\right)'+q(a)\phi(a)=\lambda \omega(a) \phi(a),
\end{equation}
where $\omega(a)$ is called the weight function, we define
\begin{equation}\left\{\begin{aligned}
&x=\int{\left(\frac{\omega(a)}{p(a)}\right)^{1/2}da},\\
&Y(x)=\left(p(a)\omega(a)\right)^{1/4}\phi(a),\\
&V(x)=\omega(a)^{-1}q(a)\\&\phantom{V(x)}-\left(\omega(a)^{-3}p(a)\right)^{1/4}\left\{p(a)\left[\left(p(a)\omega(a)\right)^{-1/4}\right]'\right\}'.
\end{aligned}\right.\end{equation}
As a result, the new function $Y(x)$ satisfies the differential equation
\begin{equation}
-Y''(x)+V(x)Y(x)=\lambda Y(x).
\end{equation}

To perform the Liouville transformation in Eq. (\ref{Sturm-Liouville}),  two distinct cases must be analyzed: $w=1$ (radiation fluid with EoS $p=\rho$) and $w\neq 1$. In each case we must be careful if $\lambda<1$ or $\lambda>1$. For $\lambda<1$, the sign of the first term in Eq. (\ref{Sturm-Liouville}) will be negative and $t=T$ will be the natural time. For $\lambda>1$, the first term in Eq. (\ref{Sturm-Liouville}) changes sign, so the time will be $t=-T$. This caution is necessary since we want to obtain a Schr\"odinger equation of the form (\ref{Sch}).

\subsection{Case $w=1$:}

We define the new variable 
\begin{equation}
x=\int{\left(\frac{\omega(a)}{p(a)}\right)^{1/2}da}=\sqrt{2m}\int{da/a}=\sqrt{2m}\ln{(a)},
\end{equation}
where $m=m_{<}\equiv 2(1-\lambda)$ if $\lambda<1$ and $m=m_{>}=2(\lambda-1)$ if $\lambda>1$. We also define the new wave function
\begin{equation}
y(x,t)=\left(p(a)\omega(a)\right)^{1/4}\psi(a,t)=\frac{1}{\left(2m\right)^{1/4}}\psi(a,t).
\end{equation}

A simple calculation shows that, in the new variable $x$, the wave function $y(x,t)$ satisfies
\begin{equation}
-\frac{\partial^2y(x,t)}{\partial x^2}+2\tilde{\Lambda} \exp{\left(\frac{2x}{\sqrt{2m}}\right)}y(x,t)=i\frac{\partial y(x,t)}{\partial t},
\label{Liouville1}
\end{equation}
where $\tilde{\Lambda}=-\Lambda$ if $\lambda<1$ and $\tilde{\Lambda}=\Lambda$ if $\lambda>1$. We see an exponential (Liouville) potential which grows with $x$ when $\tilde{\Lambda}>0$ and decreases without bound when $\tilde{\Lambda}<0 $. For $\tilde{\Lambda}=0$ we have a free particle Schr\"odinger equation.

\subsection{Case $w\neq 1$:}

In this case we define
\begin{equation}\left\{\begin{aligned}
&x=\int{\left(\frac{\omega(a)}{p(a)}\right)^{1/2}da}=\sqrt{2m}\int{da/a^{\frac{1+w}{2}}}\\&\phantom{x}=\sqrt{2m}\left(\frac{2}{1-w}\right)a^{\frac{1-w}{2}},\\
&y(x,t)=\left(p(a)\omega(a)\right)^{1/4}\psi(a,T)\\&\phantom{y(x,t)}=\frac{1}{\left(2m\right)^{1/4}}a^{\frac{1-w}{4}}\psi(a,t).
\label{transf. variables}
\end{aligned}\right.\end{equation}

A direct (tedious) calculation shows that
\begin{equation}
-\frac{\partial y(x,t)}{\partial x^2}+\left[\tilde{\Lambda}x^{2\left(\frac{1+w}{1-w}\right)}-\frac{1}{4x^2}\right]y(x,t)=i\frac{\partial y(x,t)}{\partial t},
\end{equation}
where we define
\begin{equation}\begin{aligned}
&\tilde{\Lambda}\equiv - 2\Lambda \left(2m_{<}\right)^{-\frac{1+w}{1-w}}\left(\frac{1-w}{2}\right)^{2\left(\frac{1+w}{1-w}\right)}\,\,\, \text{if} \,\,\,\lambda<1,\\
&\tilde{\Lambda}\equiv  2\Lambda \left(2m_{>}\right)^{-\frac{1+w}{1-w}}\left(\frac{1-w}{2}\right)^{2\left(\frac{1+w}{1-w}\right)}\,\,\, \text{if} \,\,\,\lambda>1.
\end{aligned}\end{equation}
Note that the resulting potential has a regular polynomial factor  $x^{2\left(\frac{1+w}{1-w}\right)}$, plus  an inverse square potential  $-\frac{1}{4x^2}$, which does not depend on the EoS.

\section{Some solvable cases}

In this section, a few cases of interest will be analyzed. In particular, the cases $w=1$ (radiation fluid), $w=0$ (dust) and $w=-1$ (dark energy) will be studied, since explicit solutions can be found in these cases. The expectation values of the scale factors in each case will be found and compared with the classical predictions of Ref. \cite{pitelli1}.

\subsection{Case $w=1$ (radiation):}
\subsubsection{$\Lambda=0$:}

In this case, the equivalent Schr\"odinger equation becomes
\begin{equation}
-\frac{\partial^2y(x,t)}{\partial x^2}=i\frac{\partial y(x,t)}{\partial t}.
\end{equation}
This is just the free particle Schr\"odinger equation. No boundary conditions are necessary at $x=\pm \infty$, since square integrability is sufficient to  uniquely  determine the solution. 

By a separation of variables $y(x,t)=X(x)e^{-i E t}$ we have $X_{k}(x)=e^{ikx}$, so that
\begin{equation}
y_k(x,t)=e^{ikx}e^{-i k^2 t},
\end{equation}
where $k^2=E$. A wave packet 
\begin{equation}
y(x,t)=\int_{-\infty}^{\infty}{y_k(x,t)A(k)dk}
\end{equation}
can then be constructed by choosing  $A(k)=e^{-\gamma k^2}$, with $\gamma>0$. This leads to
\begin{equation}\begin{aligned}
y(x,t)&=\int_{-\infty}^{\infty}{e^{i k x}e^{-i k^2 t}e^{-\gamma k^2}dk}\\&=\sqrt{\frac{\pi}{it+\gamma}}e^{-\frac{x^2}{4(it+\gamma)}}.
\end{aligned}\end{equation}
We thus have, in terms of the original variable $a$,
\begin{equation}
\Psi(a,t)=\left(2m\right)^{1/4}\sqrt{\frac{\pi}{it+\gamma}}e^{-\frac{m\left(\ln{a}\right)^2}{2(it+\gamma)}}.
\end{equation} 

\begin{figure}[h!]
\begin{center}
\includegraphics[width=8.5cm]{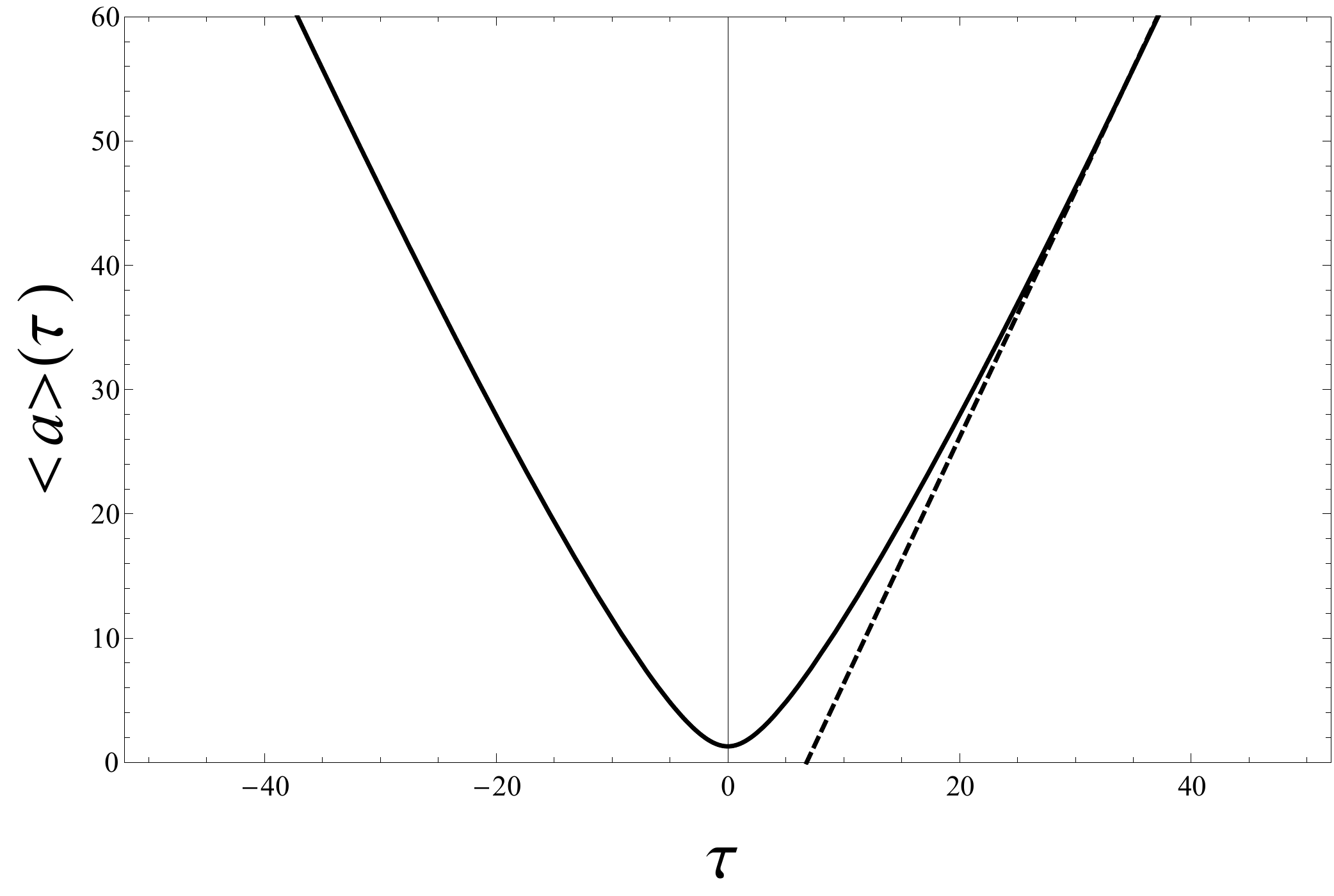}
\end{center}
\caption{Behavior of the expectation value of the scale factor as a function of the cosmic time $\tau$. We considered $\lambda=1/2$ and $\gamma=1$ in Eq. (\ref{scale factor 1}). The continuous line represents $\left<a\right>\!(\tau)$, while the dashed line represents the asymptotic behavior of $\left<a\right>\!(\tau)$.}
\label{fig1}
\end{figure}

The expectation value of the scale factor is given by
\begin{equation}
\left<a\right>\!(t)=\frac{\int_{0}^{\infty}{a\left|\Psi(a,T)\right|^2\frac{da}{a}}}{\int_{0}^{\infty}{\left|\Psi(a,t)\right|^2\frac{da}{a}}}=e^{\frac{\gamma^2+t^2}{4m\gamma}}.
\label{scale factor 1}
\end{equation}
This shows an universe bouncing from a contraction epoch to an expansion era. However, $t$ is not the cosmic time. In fact, $t$ is related to the cosmic time $\tau$ by $d\tau=N(t)dt=\left<a\right>\!(t)dt$. We see that the cosmic time cannot be expressed in terms of elementary functions. Therefore, $\tau$ is found numerically and the behavior of $\left<a\right>(\tau)$ is showed graphically in Fig. \ref{fig1}. Note that, as $\tau\to\infty$, $\left<a\right>\!(\tau)$ approaches a straight line. In Ref. \cite{pitelli1} we see that the classical solution, i.e., the solution of Eq. (\ref{equations of motion}) is given by
\begin{equation}
a(\tau)=A+B\tau.
\end{equation}
Therefore,  for asymptotically large universes we recover the classical behavior of the universe.

\subsubsection{$\Lambda\neq 0$:}

By Eq. (\ref{Liouville1}), we see that the potential of the Schr\"odinger equation in the $(x,t)$ variables has the form $V(x)=2\tilde{\Lambda}\exp{\left(\frac{2x}{\sqrt{m}}\right)}$. Since analytical solutions can only be found when the exponential potential grows with $x$, I will, therefore, confine myself to two possibilities: $\Lambda>0$ for $\lambda<1$ and $\Lambda<0$ for $\lambda>1$. In this way,  the normalized eigenstates are given by (see Ref. \cite{hoker})
\begin{equation}
\begin{aligned}
\Psi_E(a,t)=
&\left(2m\right)^{1/4}\left[\frac{2\sqrt{m}}{\pi}\sinh{\pi\left|\nu\right|}\right]^{1/2}\\&\times K_\nu\left(2\sqrt{\left|\Lambda\right| m}a\right)e^{-i E t},
\end{aligned}
\end{equation}
where $\nu=i\sqrt{2Em}$ and $K_{\nu}$ is the modified Bessel function of order $\nu$. It can be shown that the spectrum of energy is continuous and $E=0$ is a greatest lower bound. However, I was not able to find an explicit solution in this case.
\subsection{Case $w=0$ (dust):}
\subsubsection{$\Lambda=0$:}

The interesting feature with this case is that the same Schr\"odinger equation in Liouville form works for every fluid with EoS $p=w\rho$, namely,
\begin{equation}
-\frac{\partial^2y(x,t)}{\partial x^2}-\frac{1}{4x^2}y(x,t)=i\frac{\partial y(x,t)}{\partial t},
\label{-1/4}
\end{equation}
where $x=\sqrt{2m}\left(\frac{2}{1-w}\right)a^{\frac{1-w}{2}}$ and $y(x,t)=\frac{1}{\left(2m\right)^{1/4}}a^{\frac{1-w}{4}}\psi(a,t)$. With this potential, a boundary condition is necessary in order to solve Eq. (\ref{-1/4}). In Ref. \cite{gitman}, this problem has been solved and every possible boundary condition has been found. In order to obtain an explicit solution to the problem, I will choose a particular boundary condition, namely,  Friedrichs boundary condition. For this boundary condition, the energy spectrum is continuous and positive. The propagator respecting Friedrichs boundary condition can be found in Ref. \cite{efthimiou} and is given by
\begin{equation}
G(x,x';t)=\frac{1}{2t}\sqrt{x x'}i^{-1}e^{i\frac{x^2+x'^2}{4t}}J_{0}\left(\frac{x x'}{2t}\right),
\end{equation}
where $J_0$ is the $0-th$ order Bessel function.

Given an initial wave packet $y(x,0)$, we can find the solution $y(x,t)$ by means of the integral
\begin{equation}
y(x,t)=\int_{0}^{\infty}{G(x,x';t)y(x',0)dx'}.
\end{equation}
Choosing $y(x,0)=x^{1/2}e^{-\gamma x^2}$, with $\gamma>0$, we have
\begin{equation}
y(x,t)=\frac{x^{1/2}i^{-1}}{4\gamma t-i}e^{\frac{i x^2}{4t}}e^{-\frac{x^2}{16\gamma t^2-4it}}.
\end{equation}

For $w=0$ we have $x=2\sqrt{2m}a^{1/2}$ and $\psi(a,t)=(2m)^{1/4}a^{-1/4}y(x,t)$ by Eq. (\ref{transf. variables}). The wave function $\psi(a,t)$ is then given by
\begin{equation}
\psi(a,t)=\frac{\sqrt{2}\sqrt{2m}i^{-1}}{4\gamma t-i}e^{i \frac{2m a}{t}}e^{-\frac{2ma}{4\gamma t^2-i t}}.
\end{equation}
The expectation value of the scale factor can be found to be
\begin{equation}
\left<a\right>\!(t)=\frac{\int_0^{\infty}{a\left|\psi(a,t)\right|^2\frac{da}{a^{0}}}}{\int_0^{\infty}{\left|\psi(a,t)\right|^2\frac{da}{a^{0}}}}=\frac{1+16\gamma^2t^2}{16\gamma m}.
\end{equation}
Since the scale factor is given by $d\tau=Ndt$ and $N=a^{0}$, $t=\tau$ in this case. Therefore, the expectation value of the scale factor as a function of the cosmic time $\tau$ is given by
\begin{equation}
\left<a\right>\!(\tau)=\frac{1+16\gamma^2\tau^2}{16\gamma m}.
\end{equation}
The classical solution in this case (see Ref. \cite{pitelli1} and Eq. (\ref{equations of motion})) is given by
\begin{equation}
a(\tau)=A+B \tau-\frac{p_T}{2}\tau^2.
\end{equation}
Note that the quantum theory smooth out  the classical singularity, giving a bouncing model with an even expectation value of the scale factor.

\subsubsection{$\Lambda\neq 0$:}

In this case, the potential of the Schr\"odinger equation of the universe has the form $V(x)=\left[\tilde{\Lambda}x^{2}+\frac{1}{4x^2}\right]$. Analytical solutions can be found for potentials of the form $V(x)=x^2-\frac{1}{4x^2}$. In this way, I consider the cases $\Lambda>0$ if $\lambda<1$ and $\Lambda<0$ if $\lambda>1$. We arrive at the following Schr\"odinger equation
\begin{equation}
-\frac{\partial^2y(x,t)}{\partial x^2}+\left[|\tilde{\Lambda}|x^2-\frac{1}{4x^2}\right]y(x,t)=i\frac{\partial y(x,t)}{\partial t}.
\end{equation}

The propagator for the above equation can be found in \cite{efthimiou} and is given by
\begin{equation}\begin{aligned}
G(x,x';t)&=\frac{\sqrt{|\tilde{\Lambda}}|\sqrt{xx'}i^{-1}}{\sin{\left(2\sqrt{|\tilde{\Lambda}|t}\right)}}\\&\times\exp{\left[i\frac{\sqrt{|\tilde{\Lambda}|}}{2}\cot{\left(2\sqrt{|\tilde{\Lambda}|t}\right)\left(x^2+x'^2\right)}\right]}\\&\times J_{0}\left(\frac{\sqrt{|\tilde{\Lambda}|}xx'}{\sin{\left(2\sqrt{|\tilde{\Lambda}}|t\right)}}\right).
\end{aligned}\end{equation}
By choosing an initial wave packet $y(x,0)=x^{1/2}e^{-\sigma x^2}$, with $\sigma>0$, we have
\begin{widetext}
\begin{equation}\begin{aligned}
y(x,t)&=\int_0^{\infty}{G(x,x';t)y(x',0)dx'}=\frac{ i^{-1}\sqrt{|\tilde{\Lambda}|  x} }{2 \sigma \sin{\left(2\sqrt{|\tilde{\Lambda}|}  t\right)} -i \sqrt{|\tilde{\Lambda}| } \cos \left(2 \sqrt{|\tilde{\Lambda}| } t\right)}\times \\ \times&\exp \left[\frac{\sqrt{|\tilde{\Lambda}| } x^2 \left(-i \sqrt{|\tilde{\Lambda }|}+\sigma  \tan \left(\sqrt{|\tilde{\Lambda }|} t\right)-\sigma  \cot \left(\sqrt{|\tilde{\Lambda}| } t\right)\right)}{2 \sqrt{|\tilde{\Lambda}| } \cot \left(2 \sqrt{|\tilde{\Lambda} } t\right)+4 i \sigma }\right].
\end{aligned}\label{general}\end{equation}
\end{widetext}
Since $x=2\sqrt{2m}a^{1/2}$ and $\psi(a,t)=(2m)^{1/4}a^{-1/4}y(x,t)$ we have the following expression for the wave function $\psi(a,t)$ of the universe:
\begin{widetext}
\begin{equation}\begin{aligned}
\psi(a,t)&=\frac{ 2i^{-1}\sqrt{|\tilde{\Lambda}|m}  }{2 \sigma \sin{\left(2\sqrt{|\tilde{\Lambda}|}t\right)} -i \sqrt{|\tilde{\Lambda}| } \cos \left(2 \sqrt{|\tilde{\Lambda}| } t\right)}\times\\ &\times\exp \left[\frac{8m\sqrt{|\tilde{\Lambda}| } a \left(-i \sqrt{|\tilde{\Lambda }|}+\sigma  \tan \left(\sqrt{|\tilde{\Lambda }|} t\right)-\sigma  \cot \left(\sqrt{|\tilde{\Lambda}| } t\right)\right)}{2 \sqrt{|\tilde{\Lambda}| } \cot \left(2 \sqrt{|\tilde{\Lambda} } t\right)+4 i \sigma }\right].
\end{aligned}\end{equation}
\end{widetext}
The expectation value of the scale factor as a function of the cosmic time $\tau$ is then given by
\begin{equation}
\left<a\right>\!(\tau)=\frac{4\sigma^2\sin^2{\left(2\sqrt{|\tilde{\Lambda}|}\tau\right)}+|\tilde{\Lambda}|\cos^2{\left(2\sqrt{|\tilde{\Lambda}|}\tau\right)}}{16m|\tilde{\Lambda}|\sigma}.
\end{equation}
This shows an oscillatory behavior as predicted by the classical theory \cite{pitelli1}. Once again the big-bang singularity is smoothed and the expectation value of the scale factor is an even function of the cosmic time.

\subsection{Case $w=-1$ (dark energy):}
\subsubsection{$\Lambda=0$:}
In this case $x=\sqrt{2m}a$ and $\psi(a,t)=(2m)^{1/4}a^{-1/2}y(x,t)$ by Eq. (\ref{transf. variables}).  By Eq. (\ref{general}), the wave function $\psi(a,t)$ becomes
\begin{equation}
\psi(a,t)=\frac{\sqrt{2m}i^{-1}}{4\gamma t -i}e^{i\frac{m a^2}{2t}}e^{-\frac{2m a^2}{16 \gamma t^2-4i t}}.
\end{equation}
The expectation value of the scale factor is given by 
\begin{equation}
\left<a\right>\!(t)=\frac{\int_0^{\infty}{a\left|\psi(a,t)\right|^2\frac{da}{a^{-1}}}}{\int_0^{\infty}{\left|\psi(a,t)\right|^2\frac{da}{a^{-1}}}}=\frac{\sqrt{\pi}}{4\sqrt{m\gamma}}\sqrt{1+16\gamma^2t^2}.
\end{equation}
The proper time $\tau$ can be found from $d\tau=a^{-1}dt$. It is given by
\begin{equation}
\tau=\sqrt{\frac{m}{\pi \gamma}}\text{arcsinh}\left(4\gamma t\right).
\end{equation}
Therefore, the expectation value of the scale factor as a function of the cosmic time has the form
\begin{equation}
\left<a\right>\!(\tau)=\frac{\sqrt{\pi}}{4\sqrt{m\gamma}}\cosh{\left(\sqrt{\frac{\pi \gamma}{m}}\tau\right)}.
\label{w=-1}
\end{equation}
This inflationary universe is what we expect from a fluid with equation of state $p=-\rho$.
\subsubsection{$\Lambda\neq 0$:}

In this case we have
\begin{equation}
-\frac{\partial y(x,t)}{\partial x^2}+\left[\tilde{\Lambda}-\frac{1}{4x^2}\right]y(x,t)=i\frac{\partial y(x,t)}{\partial t},
\end{equation}
The only effect of $\tilde{\Lambda}$ is to shift the energy, i.e., if $y(x,t)$ is a solution of the above equation with $\tilde{\Lambda}=0$, then $y(x,t)e^{-i \tilde{\Lambda}t}$ is a solution of the above equation with $\tilde{\Lambda}\neq0$. Therefore, the scale factor does not change and it is still given by Eq. (\ref{w=-1}).

\acknowledgments

This work was partially supported by FAPESP Grant. No. 2013/09357-9.

\section{Conclusions}

In this paper, the $(1+1)$-dimensional FRW universe filled with a perfect fluid in HL theory has been quantized. The resulting Schr\"odinger equation of the universe has some interesting properties. For a radiation fluid, it corresponds to a free particle equation when the cosmological constant is null and has a Liouville potential when $\Lambda\neq 0$. For all other kinds of fluids, the Schr\"odinger equation have a potential composed by a characteristic inverse square potential, not depending on the EoS, plus a regular polynomial (when $\Lambda\neq 0$) which depends on the equation of state. 

A few solvable cases were studied and the evolution of the scale factor was found. Consistency with the classical predictions found in Ref. \cite{pitelli1} was proved.  In all cases, the universe bounces around the big-bang singularity $a=0$ and tends to the classical universe when the cosmological time $\tau$ goes to infinity. This also happens in usual quantum cosmology \cite{alvarenga}. In this way, in GR and in HL theory, quantization seems to smooth out the big-bang singularity at $a=0$, while still retaining the classical behavior as the universe becomes asymptotically large.


\begin{thebibliography}{99}
\bibitem{horava1}  P. Ho\v rava, Phys. Rev. D {\bf 79}, 084008 (2009).

\bibitem{visser1}  M. Visser, Phys. Rev. D {\bf 80}, 025011 (2009).

\bibitem{wang2}
A. Wang, D. Wands and R. Maartens, JCAP {\bf 1003}, 013 (2010).
\bibitem{kiritsis} 
E. Kiritsis and G. Kofinas,  Nucl. Phys. B {\bf 821},467 (2009).
\bibitem{mukohyama} 
S. Mukohyama,   JCAP {\bf 0906}, 001 (2009).

\bibitem{mukohyama2} 
S. Mukohyama, Class. Quant. Grav. {\bf 27}, 223101 (2010).



\bibitem{wang}
Bao-Fei Li,  A. Wang, Y. Wu and Z. C. Wu, Phys. Rev. D {\bf 90}, 124076 (2014).

\bibitem{pitelli1}
J.P.M. Pitelli, Phys. Rev. D, {\bf 92}, 084012 (2015).




\bibitem{grumiller}
D. Grumiller, W. Kummer and D.V. Vassilevich, Phys. Rept. {\bf 369}, 327 (2002).

\bibitem{cadoni} M. Cadoni, S. Mignemi, Gen. Rel. Grav. {\bf 34}, 2101 (2002).





\bibitem{dewitt}
B. DeWitt, Phys. Rev. {\bf 160} (1967) 1113.

\bibitem{adm} R.L. Arnowitt, S. Deser, and C.W. Misner, {\it The dynamics of general relativity, in Gravitation: An Introduction to Current
Research}, edited by L. Witten, John Wiley, New York, (1962).

\bibitem{schutz1} B.F. Schutz, Phys. Rev. D {\bf 2}, 2762 (1970).

\bibitem{schutz2}  B.F. Schutz, Phys. Rev. D {\bf 4}, 3559 (1971).

\bibitem{pitelli2}
J.P.M. Pitelli and A. Saa, Phys. Rev. D {\bf 86}, 063506 (2012).
\bibitem{vakili}
B. Vakili and V. Kord, Gen. Rel. Grav. {\bf 45}  1313 (2013).

\bibitem{reed}
M. Reed and B. Simon, {\it Fourier Analysis and Self-Adjointness}, Academic Press, New York, (1972).

\bibitem{ahmed}
M.A. Ahmed, Phys. Rev. D {\bf 61}, 104012 (2000).

\bibitem{Lapchinskii}
V.G. Lapchinskii and V.A. Rubakov, Theor. Math. Phys. {\bf 33}, 1076 (1977).

\bibitem{dirac}
P.A.M. Dirac,{\it Lectures on Quantum Mechanics}, Dover Publications, Inc., New York, (2001).


\bibitem{everitt}

W.N. Everitt, {\it  A Catalogue of Sturm-Liouville Differential Equations. Sturm-Liouville theory},
pp. 271–331. Birkhäuser, Basel (2005)

\bibitem{hoker}

E. D'Hoker and R. Jackiw, Phys. Rev. D {\bf 26}, 3517 (1982).

\bibitem{gitman}
D.M. Gitman, I.V. Tyutin and B.L. Voronov, J. Phys. A: Math. Theor. {\bf 43}, 145205 (2010).

\bibitem{efthimiou}
C.J. Efthimiou, Phys. Rev. A, {\bf 48}, 4758 (1993).

\bibitem{alvarenga}
F.G. Alvarenga, J.C. Fabris, N.A. Lemos and G.A. Monerat, Gen. Rel. Grav. \bf{34}, 651 (2002).
\end{thebibliography}
\end{document}